\documentclass[preprint,longabstract]{aastex}

\usepackage{rotating}
\usepackage{graphicx}

\newcommand{\e}{\epsilon}
\newcommand{\g}{\gamma}
\newcommand{\gp}{\gamma^{\prime}}

\newcommand{\psim}{\lower.5ex\hbox{$\; \buildrel \propto \over\sim \;$}}
\newcommand{\lbar}{\lower.0ex\hbox{$\; \buildrel
{\lower0.0ex \hbox{-}} \over\lambda  \;$}}

\newcommand{\cm}{\mathrm{cm}}

\newcommand{\erg}{\mathrm{erg}}

\newcommand{\s}{\mathrm{s}}

\newcommand{\LAT}{\textit{Fermi}-LAT }

\slugcomment{To appear in the Astrophysical Journal}

\shorttitle{Multiband Observations of PKS\,2326$-$502}
\shortauthors{Dutka, Ojha, D'Ammando et al.}

\begin{document}


\title{Multiband Observations of the Quasar PKS\,2326$-$502 during Active and Quiescent Gamma-Ray States in 2010-2012}

\author{Michael S.\ Dutka\altaffilmark{3}}
\affil{The Catholic University of America, 620 Michigan Ave., N.E.  Washington, DC 20064}
\email{ditko86@gmail.com}

\author{Bryce D.\ Carpenter\altaffilmark{2}}
\affil{NASA Goddard Space Flight Center, Astrophysics Science Division, Code 661, Greenbelt, MD 20771}
\email{carpbr01@gmail.com}

\author{Roopesh Ojha\altaffilmark{1}}
\affil{UMBC/NASA Goddard Space Flight Center, Astrophysics Science Division, Code 661, Greenbelt, MD 20771}
\email{roopesh.ojha@gmail.com}

\author{Justin D.\ Finke}
\affil{Naval Research Laboratory, Space Science Division, Code 7653, 4555 Overlook Ave. SW, Washington, DC 20375}

\author{Filippo D'Ammando}
\affil{Universit\`a di Bologna Dipartimento di Fisica e Astronomia, INAF-IRA, Bologna, Italy}

\author{Matthias Kadler}
\affil{Lehrstuhl f\"{u}r Astronomie, Universit\"{a}t W\"{u}rzburg, Emil
-Fischer-Stra{\ss}e 31, D-97074 W\"urzburg, Germany}

\author{Philip G.\ Edwards}
\affil{CSIRO Astronomy and Space Science, PO Box 76, Epping NSW 1710, Australia}

\author{Jamie Stevens}
\affil{CSIRO Astronomy and Space Science, 1828 Yarrie Lake Road, Narrabri NSW 2390, Australia}


\author{Eleonora Torresi}
\affil{Istituto Nazionale di Astrofisica, (National Institute of Astrophysics) INAF-IASFBO, via Gobetti 101, I-40129 Bologna, Italia}

\author{Paola Grandi}
\affil{Istituto Nazionale di Astrofisica, (National Institute of Astrophysics) INAF-IASFBO, via Gobetti 101, I-40129 Bologna, Italia}

\author{Roberto Nesci}
\affil{Istituto Nazionale di Astrofisica, (National Institute of Astrophysics) INAF-IAPS, via Fosso del Cavaliere 100, 00133 Roma, Italia}

\author{Felicia Krau\ss}
\affil{GRAPPA \& Anton Pannekoek Institute for Astronomy, University of Amsterdam, Science Park 904, 1098 XH Amsterdam, The Netherlands}

\author{Cornelia M\"uller}
\affil{Department of Astrophysics/IMAPP, Radboud University Nijmegen, PO Box 9010, 6500 GL, Nijmegen, The Netherlands}

\author{Joern Wilms}
\affil{Remeis Observatory \& ECAP, Sternwartstr. 7, 96049 Bamberg, Germany}

\author{Neil Gehrels}
\affil{NASA Goddard Space Flight Center, Astrophysics Science Division, Code 661, Greenbelt, MD 20771}


\altaffiltext{1}{Adjunct Professor, The Catholic University of America, 620 Michigan Ave., N.E.  Washington, DC 20064}
\altaffiltext{2}{The Catholic University of America, 620 Michigan Ave., N.E.  Washington, DC 20064}
\altaffiltext{3}{Wyle Science Technology and Engineering group, NASA GSFC, Greenbelt, MD 20771}


\begin{abstract}
Quasi-simultaneous observations of the Flat Spectrum Radio Quasar PKS\,2326$-$502 were carried out in the $\gamma$-ray, X-ray, UV, optical, near-infrared, and radio bands. Thanks to these observations we are able to characterize the spectral energy distribution of the source during two flaring and one quiescent $\gamma$-ray states. These data were used to constrain one-zone leptonic models of the spectral energy distributions of each flare and investigate the physical conditions giving rise to them. While modeling one flare only required changes to the electron spectrum, the other flare needed changes in both the electron spectrum and the size of the emitting region with respect to the quiescent state. These results are consistent with an emerging pattern of two broad classes of flaring states seen in blazars. 

\end{abstract}


\section{Introduction}\label{introduction}

PKS\,2326$-$502 is a Flat Spectrum Radio Quasar (FSRQ) at a tentative (based on a single emission line) redshift of 0.518 \citep{Jauncey1984}. It is present in all three \textit{Fermi} Large Area Telescope (\textit{Fermi}-LAT) catalogs (1FGL: \citet{Abdo2010_1fgl}; 2FGL: \citet{Nolan2012}; 3FGL: \citet{3FGL}). In the 3FGL it is listed as 3FGL\,J2329.3$-$4955 with a (0.1 --100) GeV flux of (25.1 $\pm$ 0.4) $\times \ 10^{-8}$ ph cm$^{-2}$ s$^{-1}$ averaged over the first four years of the \textit{Fermi} mission. PKS\,2326$-$502 is one of the southern hemisphere $\gamma$-ray loud active galactic nuclei (AGN) being studied by the Tracking Active Galactic Nuclei with Austral Milliarcsecond Interferometry (TANAMI) program \citep{Ojha2010} at radio and other wavelengths.

Typically divided into BL Lacertae objects and FSRQs, blazars are the most luminous subclass of radio loud  AGN.  
They commonly exhibit high polarization levels and variability on a wide range of timescales and in every wavelength. Blazars often show collimated structures called ÔjetsÕ that are thought to be powered by the accretion of material onto a spinning black hole \citep{Blandford1977}. Jets, which can appear to be moving at superluminal speeds, are ubiquitous among radio loud AGN and can extend for thousands of parsecs from the supermassive black hole at the center of their host galaxy. The high luminosity of blazars is a consequence of the relativistic Doppler boosting that results from the small angles between their jets and the observer's line of sight \citep{Urry1995}.   For this reason blazars are the largest subclass of objects detected by the large area telescope on the
 \emph{Fermi Gamma-ray Space Telescope} \citep{3LAC}.
 
The broad band spectral energy distribution (SED) for most blazars has a roughly `double-humped'
 shape, with low and high-energy components. Blazars can be classified based on the frequency of 
the peak of the low-energy component in a $\nu F_{\nu}$ representation.  Evidence suggests that the low-energy component is caused by synchrotron emission \citep{Urry1995}, therefore we refer to this peak as the synchrotron peak.  Low-synchrotron peaked (LSP) blazars have $\nu^{\textnormal{\small sync}}_{\textnormal{\small peak}}$ at $<10^{14}$\ Hz; high-synchrotron peaked (HSP) blazars have $\nu^{\textnormal{\small sync}}_{\textnormal{\small peak}}$ $>10^{15}$\ Hz; and intermediate-synchrotron peaked (ISP) blazars have $\nu^{\textnormal{\small sync}}_{\textnormal{\small peak}}$ between $10^{14}$ and $10^{15}$\ Hz \citep{Abdo2010_sed}. Almost all FSRQs, of which PKS\,2326$-$502 is an example, are LSP blazars \citep{Finke2013ApJ763134}.  

The emission mechanisms responsible for the high-energy component of the blazar SED are not very well established.  There are two broad classes of models for its origin, hadronic and leptonic \citep{Bottcher2007}.  Hadronic models assume that a large fraction of the jet power goes into the acceleration of protons, and then these protons and their secondaries are responsible for the high-energy emission \citep{Mannheim1992}.  Leptonic models invoke inverse Compton scattering of seed photons by the electrons in the jet. Leptonic models can be subdivided into many different types depending on the source of seed photons and the number of emitting zones \citep{Bottcher2007}.  Some models invoke large numbers of emitting zones \citep{Marscher2014} while others invoke only a single (or a few) emitting regions \citep{Dermer2009}.  Leptonic models that use the synchrotron photons already present in the jet as the source of seed photons for inverse Compton scattering are referred to as synchrotron self Compton (SSC) models \citep{Maraschi2003}. Models that use photons originating outside the jet (e.g., the accretion disk or the dust torus) are known as external Compton (EC) models \citep{Dermer2009}. Due to the variability of blazars in every observing band, differentiating between these physical scenarios is only possible with simultaneous or close to simultaneous multiwavelength (MWL) observations in different activity states.

On 2010 August 7, the \textit{Fermi}-LAT detected PKS\,2326$-$502 at a daily averaged flux above 100 MeV of (1.1 $\pm$ 0.3) $\times \ 10^{-6}$ ph cm$^{-2}$ s$^{-1}$, a factor of 15 brighter than the average over the first eleven months of \textit{Fermi}-LAT operations \citep{ATel2783} and a factor of 4.4 brighter than the 3FGL average flux. Follow-up X-ray, UV and optical observations were made by the \textit{Swift} XRT and UVOT on 2010 August 18. Two years later, on 2012 June 27, PKS\,2326$-$502 was detected with a daily averaged flux of (1.4 $\pm$ 0.3) $\times \ 10^{-6}$ ph cm$^{-2}$ s$^{-1}$ by the \textit{Fermi}-LAT. This was a factor of 11 greater than the average flux over the first two years of the mission \citep{ATel4225} and a factor of 5.6 greater then the 3FGL flux. Again \textit{Swift} made follow up observations on 2012 June 29  providing X-ray, UV and optical measurements. Additional observations were made in the optical band by the ANDICAM at Cerro Tololo on 2012 June 30 and the Rapid Eye Mount  0.6 m telescope in La Silla, Chile on 2012 July 01 and 02. Radio measurements were made by the Australia Telescope Compact Array on 2012 June 29 as part of the TANAMI monitoring program.

In order to investigate the high-energy emission from PKS\,2326$-$502, we have defined two flaring and one quiescent $\gamma$-ray states based on the \textit{Fermi}-LAT light curve (Figure \ref{gammalightcurve}). The first flare lasted from 2010 July 31 to 2010 September 29 (hereafter Flare `A') and the second from 2012 June 25 to 2012 July 05 (hereafter Flare `B'). In order to provide a baseline to compare the flaring states, observations that were performed during a $\gamma$-ray quiescent state from 2011 December 18 to 2012 January 29 (hereafter period 'Q') are also considered. 
 
This paper is organized as follows. In section~\ref{gamma} we discuss the LAT analysis, followed by sections \ref{swift} -- \ref{radio} describing observations at X-ray, ultraviolet, optical and radio wavelengths respectively. In section \ref{results} we describe our modeling and discuss what our observations and modeling suggest. The conclusions can be found in section  \ref{conclusions}. We assume H$_{0}$ = 70 km s$^{-1}$ Mpc$^{-1}$, $\Omega_{m}$ = 0.3 and $\Omega_{\lambda}$ = 0.7.

\section{LAT Observations}\label{gamma}

The \textit{Fermi}-LAT \citep{Atwood2009} is one of the two instruments onboard the \textit{Fermi Gamma-ray Space Telescope} . It is a $\gamma$-ray pair production detector providing unprecedented all-sky spatial and energy resolution in the 100 MeV -- 300 GeV band. It typically operates in an all-sky survey mode and its 2.5 steradian field of view allows it to monitor the entire sky once every 3 hours, enabling rapid response to extraordinary $\gamma$-ray flaring activity. Thus \textit{Fermi}-LAT is an ideal instrument to trigger near simultaneous broadband coverage. 

A light curve has been created from 2009 February 21 (MJD 54883) to 2012 December 5 (MJD 56266) in order to determine the duration of the flaring periods and select a quiescent state. For each of these states spectral fitting was used to determine the flux in multiple energy bands.

\begin{figure}
\plotone{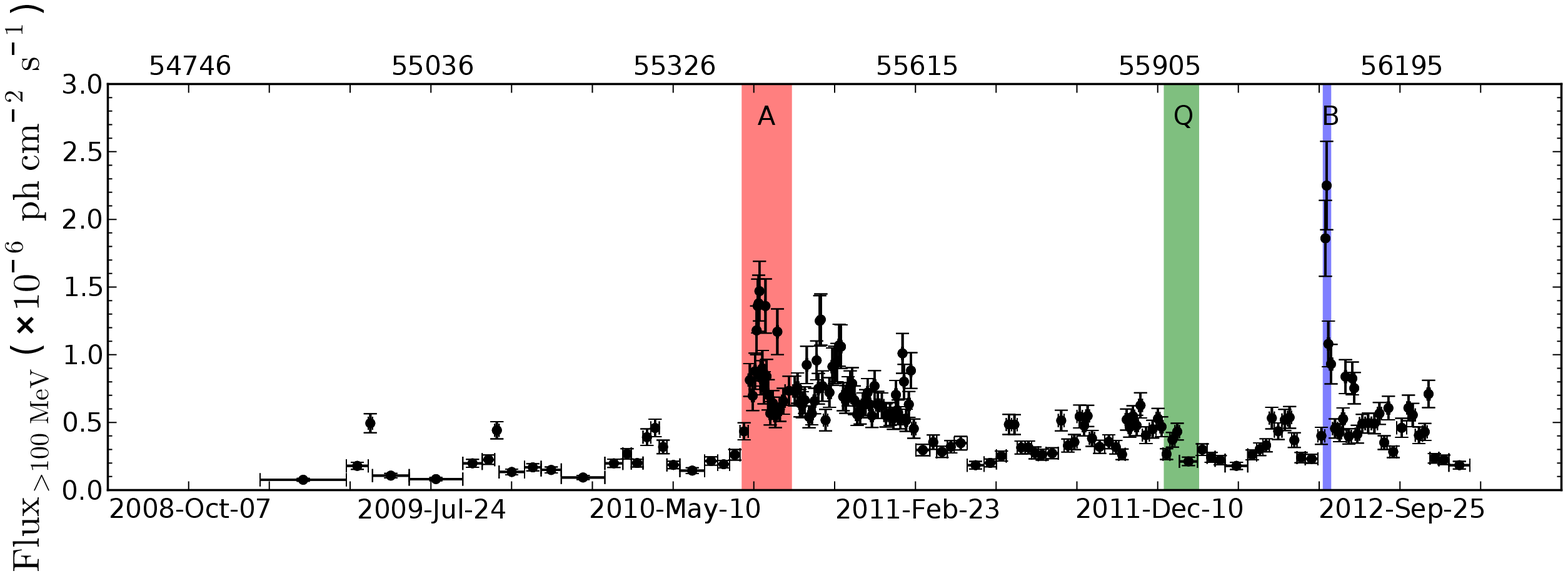}
\caption{{\em Fermi}-LAT adaptively binned $\gamma$-ray light curve in the 100 MeV to 300 GeV energy range for PKS\,2326$-$502. It covers the period from 2009 February 21 (MJD 54883) to 2012 December 05 (MJD 56266). The bin sizes are set such that a constant flux uncertainty of 15\% is maintained. Three different states of the source are selected and their MWL SEDs are modeled. The long active state is from 2010 July 31 - 2010 September 29 (flare A, in red), the quiescent state from 2011 December 18 through 2012 January 29 (period Q, in green), and a short, high peaked, flaring state from 2012 June 25 through 2012 July 05 (flare B, in purple).
\label{gammalightcurve}}
\end{figure}

The light curve, shown in Figure \ref{gammalightcurve}, was created with data in the 100 MeV to 300 GeV energy range using the adaptive binning approach described in \citet{Lott2012}. A caveat with this method is that fluxes for nearby point sources are not accounted for. However PKS\,2326$-$502 is well outside the Galactic plane and has no bright nearby sources (nearest source is 1.2\arcdeg, the next closest more the 2\arcdeg\ away) which makes it an excellent source for the adaptive binning method. The adaptive binning method varies the time bin size so that each bin has a fixed flux error of 15\%. The light curves show doubling times of $(4.5\ \times\ 10^{4}$ s during flare A and  $6.2\ \times\ 10^{4}$ s during flare B.

For both the light curve and spectral fitting, events above a zenith angle of 100\arcdeg \ were cut and a rocking angle cut of 52\arcdeg\ was applied to avoid contamination from the Earth limb. The Galactic diffuse emission and the isotropic background were accounted for using the models gal\_2yearp7v6\_v0.fits and iso\_p7v6source.txt\footnote{http://fermi.gsfc.nasa.gov/ssc/data/access/lat/BackgroundModels.html} with fixed normalizations. The analysis was done using Fermi science tools version 09-27-00 with instrument response function P7SOURCE\_V6.

LAT spectral analysis was conducted on two flaring states (flare A and flare B), and for the quiescent period Q. The \lq source' event class was selected and data extracted from a circular region of interest (ROI) of 10\arcdeg\ centered on PKS\,2326$-$502.  The starting spectral model of the source used a power law with a spectral index of 2.24 and a flux of $1.2 \times10^{-7}$ ph cm$^{-2}$ s$^{-1}$. These starting parameters are the averages from the Fermi 2FGL catalog. These parameters were refitted for each period based on the data. The model of the ROI contains the 2FGL information on all sources within 20\arcdeg\ of the source. Sources within 10\arcdeg\ had the normalization parameter free and the index fixed to values determined by a likelihood analysis across the entire energy range during each period. For those sources outside 10\arcdeg\ both parameters were fixed to the 2FGL values. The spectral indices for other sources were held fixed for the determination of the spectral points and the lightcurve. The fluxes were determined by likelihood analyses using the \texttt{gtlike} tool of the Fermi Science Tools.

Analysis was conducted using several different spectral shapes for the $\g$-ray data (power law, 
broken power law, and log parabola)\footnote{http://fermi.gsfc.nasa.gov/ssc/data/analysis/scitools/xml\_model\_defs.html}.
Flare A  was best fit by the power law model, showing no significant curvature. In flare B the power law fit was slightly worse than the other models, with a preference for the log parabola over the power law of 3 $\sigma$. The quiet state showed no significant preference, with differences of between 1 and 2 $\sigma$ between the models significance.

The two flaring states differ in length: flare A showed increased emission for an extended period of time as can be seen in Figure \ref{gammalightcurve}, whereas flare B showed a sharp peak after which the flux very quickly returned to its previous state. Flare A had an average spectral index of $2.23 \pm 0.03$, a peak flux of $(1.47\pm0.22) \times10^{-6}$ ph cm$^{-2}$ s$^{-1}$ and average flux of $(8.82\pm1.32) \times10^{-7}$ ph cm$^{-2}$ s$^{-1}$. During flare B the average spectral index was $1.99 \pm 0.05$ with peak flux $(2.25\pm0.33) \times10^{-6}$ ph cm$^{-2}$ s$^{-1}$ and average flux $(1.53\pm0.23) \times10^{-6}$ ph cm$^{-2}$ s$^{-1}$. Period Q had a spectral index of $2.25 \pm 0.06$ and average flux $(3.2\pm0.5) \times10^{-7}$ ph cm$^{-2}$ s$^{-1}$. To determine individual spectral points, analysis was run on energy bins using the spectral index found by analysis over the entire LAT energy spectrum. These spectra can be seen in Figure 2. The size of the bins varies due to the photon statistics available for each period.

\begin{figure}
\includegraphics[width=0.50\textwidth]{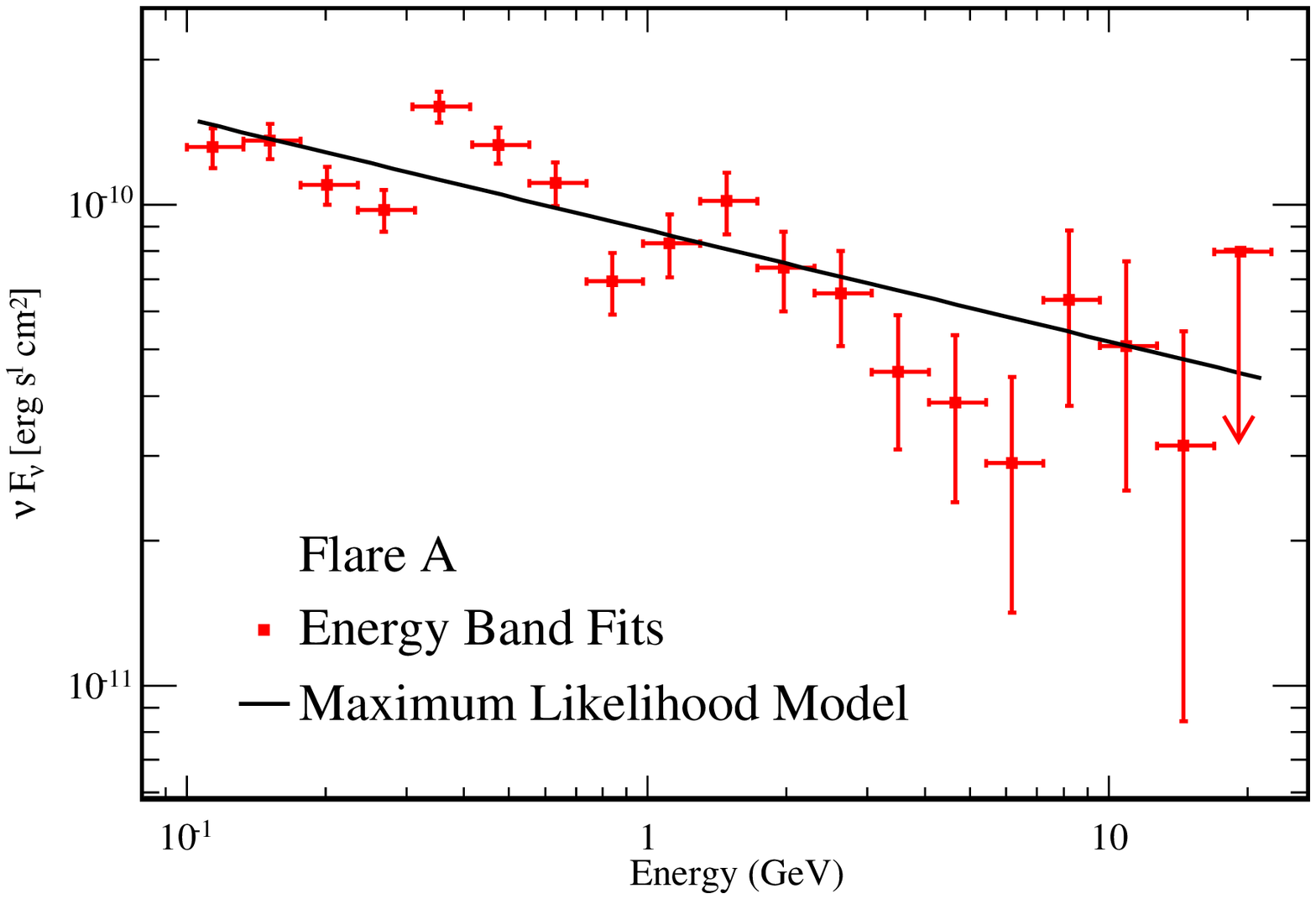}\hfill
\includegraphics[width=0.50\textwidth]{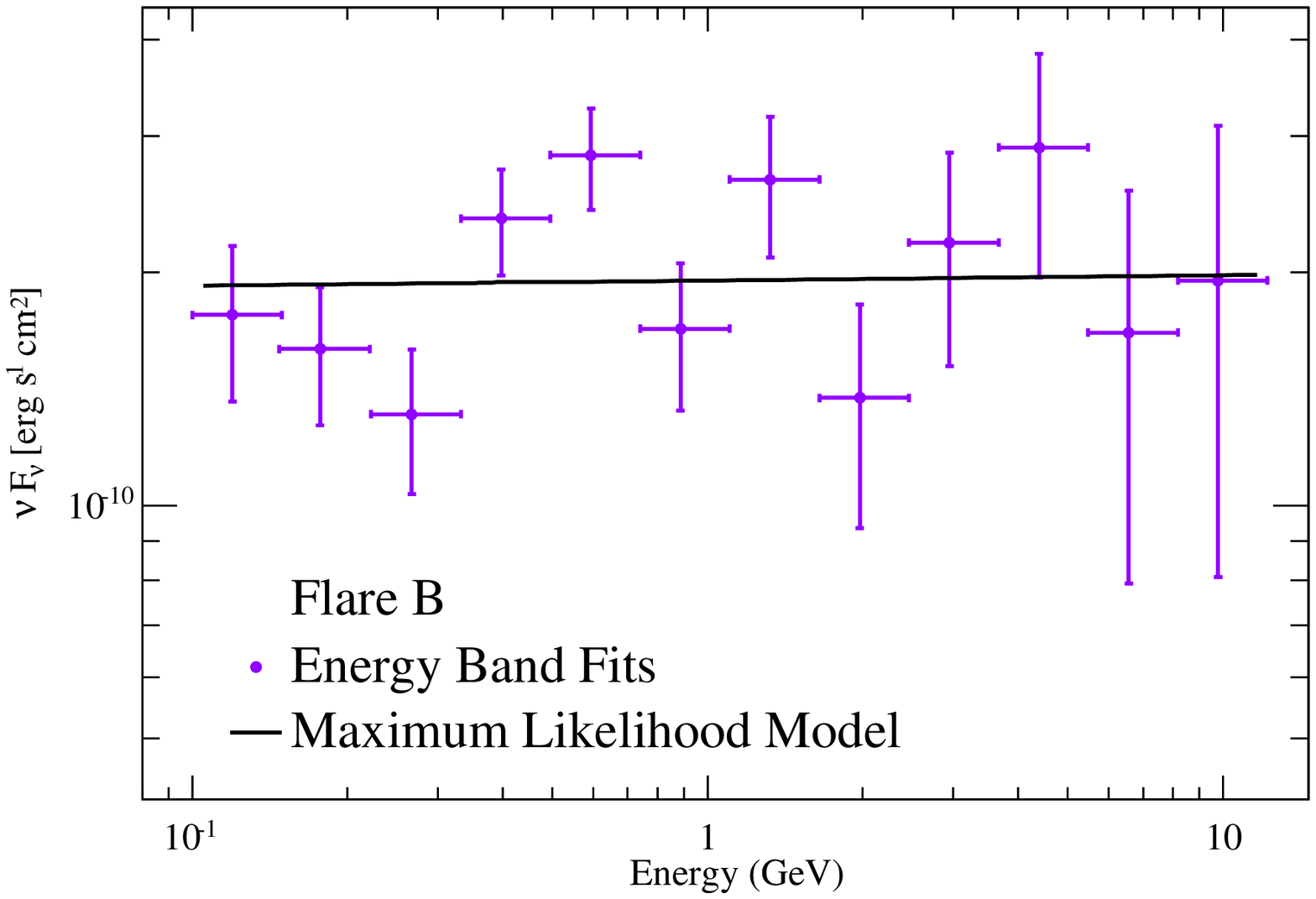}\hfill
\includegraphics[width=0.50\textwidth]{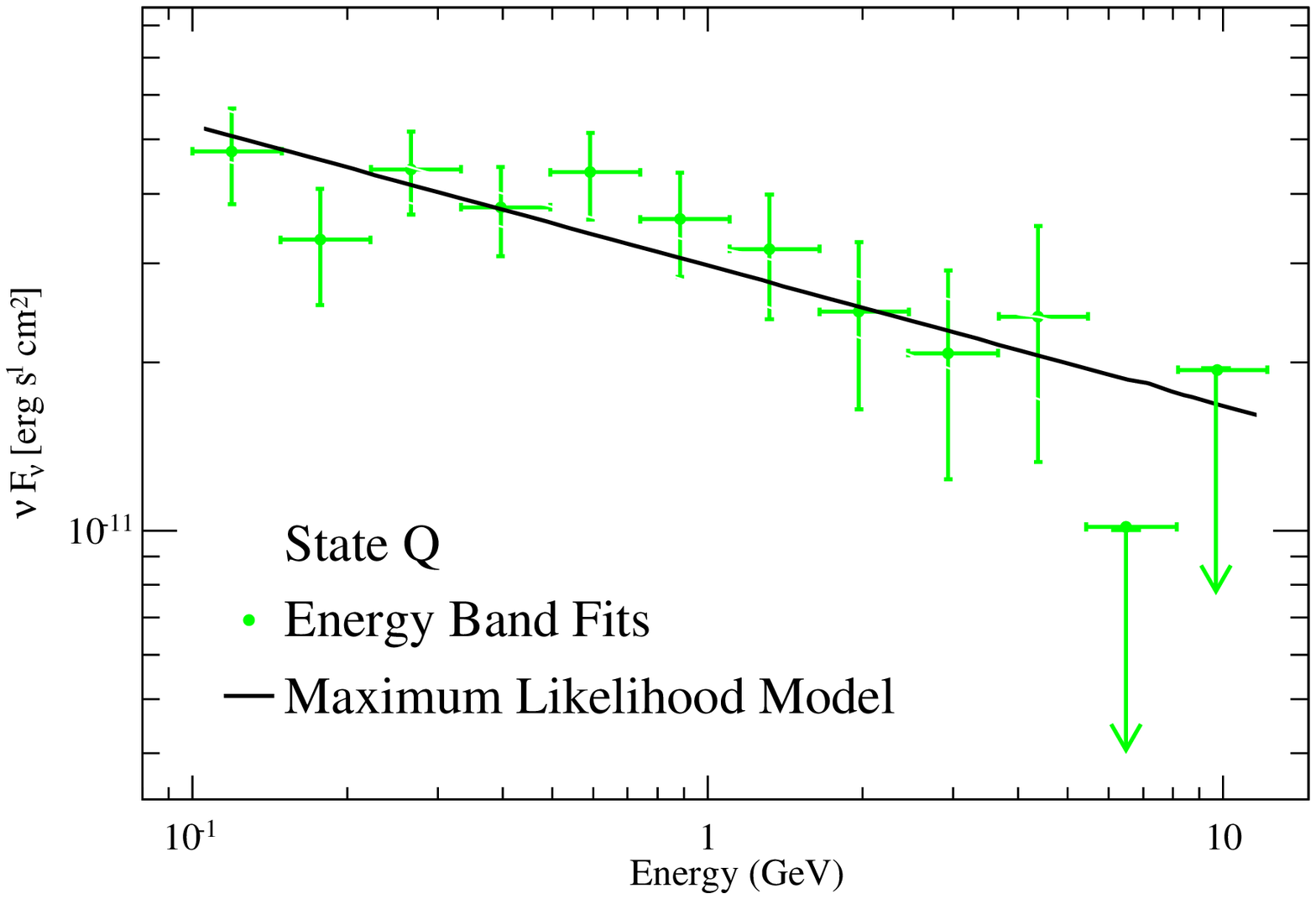}\hfill
\includegraphics[trim=0 0 -80 0,clip,width=0.5\textwidth]{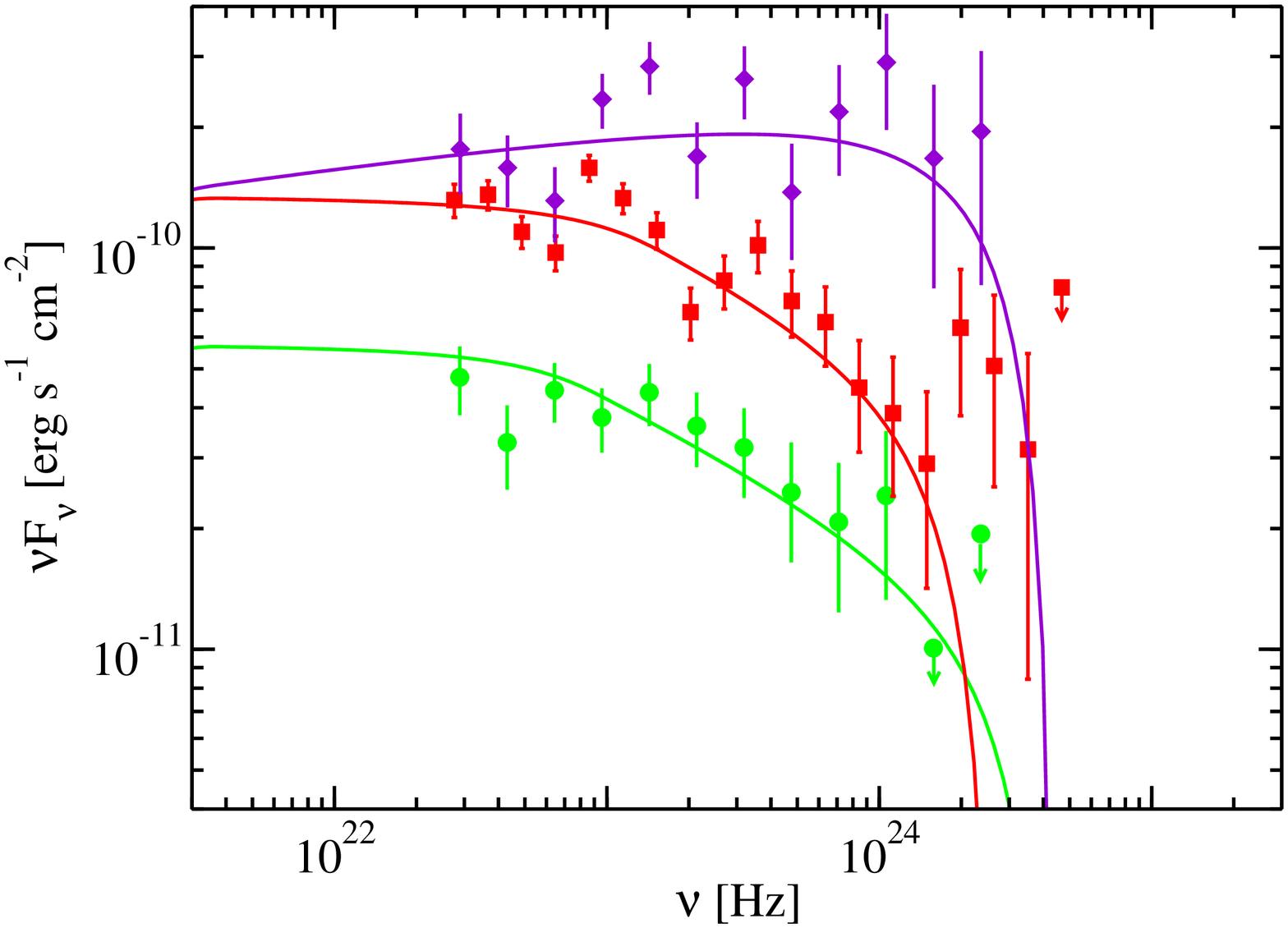}
\caption{$\gamma$-ray SEDs of PKS\,2326$-$502. The top left plot shows the SED during flare A. Top right is flare B. Bottom left is period Q. The black line on each plot shows the power-law fit. The bottom right plot shows all three states plotted together. The green, red and purple points represent the quiet, A, and B states respectively. Note that the solid lines on the bottom right figure are the SED models, not fits to the data.
\label{gammaspectra}}
\end{figure}

\section{{\em Swift} Observations}\label{swift}
The \textit{Swift} observatory \citep{Gehrels2004} was designed as a rapid response mission for expeditious follow-up of $\gamma$-ray bursts. Its quick slew rate and the availability of Target of Opportunity observations make \textit{Swift} a powerful tool for the study of AGN flares detected by the \textit{Fermi}-LAT. During this observing campaign we used the data from two of the instruments on board \textit{Swift}, the UltraViolet and Optical Telescope (UVOT), and the X-Ray Telescope (XRT).

Observations of PKS\,2326$-$502 were made during both flares and the quiescent state, on 2010 August 18,  2011 December 30 and 31, and 2012 June 29, respectively. During flare A the XRT observed for 4.7 ks and UVOT observations were made with the V, B, U, W1, M2, and W2 filters. For the period Q the XRT took 2.7 ks of data and the UVOT observed with the W1, M2, and W2 filters. During flare B the XRT observed for 2.0 ks and UVOT observed with the U and W2 filters.

The XRT spectrum was fit in the 0.3--10 keV range and the fit used an 
absorbed power law with an $N_{H}$ column density of $1.18 \times 10^{20}\ $cm$^{-2}$ \citep{Kalberla2005}.
The XRT data were processed by using the \texttt{xrtpipeline} of the
\texttt{HEASoft} \footnote{http://heasarc.nasa.gov/lheasoft/} package (v6.14) with standard procedures, filtering, and
screening criteria. Considering the low number of photons collected ($<$
200 counts) the spectra were rebinned with a minimum of 1 count per bin
and we performed the fit with the Cash statistic \citep{cash79}. Source
events were extracted from a circular region with a radius of 20 pixels (1
pixel $\sim$ 2.36 arcsec), while background events were extracted from a
circular region with radius of 50 pixels far away from bright sources.
Ancillary response files were generated with \texttt{xrtmkarf}, and
account for different extraction regions, vignetting and point spread
function corrections. We used the spectral redistribution matrices in the
calibration database maintained by HEASARC\footnote{http://heasarc.nasa.gov/}.

UVOT data were analyzed with the \texttt{uvotsource} task included in the
\texttt{HEASoft} package (v6.14). Source counts were extracted from a
circular region of 5 arcsec radius centered on the source, while
background counts were derived from a circular region with 10 arcsec
radius in a nearby source-free region. The central wavelengths of the filters are V: 5468 \AA, B: 4392 \AA, U: 3465 \AA, UVW1: 2600 \AA, UVM2: 2246 \AA, and UVW2: 1928 \AA. Galactic extinction was corrected for using the method from \citet{Fitzpatrick1999} and the method described in \citet{Predehl1995} was used to calculate the extinction parameter $E(B-V)$ from the $N_{H}$ column density. The $N_{H}$ Column density is 1.18 $\times$ 10$^{20}$ cm$^{-2}$. The relation from \citet{Predehl1995} is $E(B-V) = 5.3\ \times\ 10^{21}\ N_{H}$ giving an $E(B-V)$ value of 0.022. 



\section{Optical/NIR Observations}

Regular observations of many \LAT and TANAMI blazars are made by the Small and Moderate Aperture
Research Telescope System \citep[SMARTS;][]{Bonning2012}. Providing optical and IR photometric data, SMARTS uses the ANDICAM mounted on the 1.3\,m telescope located at the Cerro Tololo Inter-American Observatory.  The ANDICAM uses a dichroic to take simultaneous optical and infrared data with a CCD and a HgCdTe array. The IR exposures can be dithered during the optical exposure through the use of a  moveable mirror. SMARTS observed PKS\,2326$-$502 on 2012 June 30 in the $R$ band contemporaneously to flare B. The other two periods were not observed by SMARTS. 

The 0.6\,m Rapid Eye Mount \citep[REM;][]{REM2003} telescope is primarily designed to provide rapid response to $\gamma$-ray bursts detected by \textsl{Swift} and other satellites. It is located on the La Silla premises of the ESO Chilean Observatory. REM observed PKS\,2326$-$502 in the J and K bands on 2012 July 01 and the H band on 2012 July 02 contemporaneously to flare B. Photometric data from REM were analysed using the IRAF/Apphot package\footnote{ftp://iraf.noao.edu/ftp/docs/apuser.ps.Z}. Photometric measurements were made on the source as well as several nearby stars within 5 arcmin surrounding the source in the sky. A linear fit between the instrumental magnitudes of these stars and their catalog magnitudes was used to find the magnitude of PKS\,2326$-$502. Error estimates were obtained from the root mean square deviation of the reference stars from the best fit line. 

These observations were all corrected for Galactic extinction \citep{Schlafly2011} and converted from the magnitude system to fluxes using photometric zero points from \citet{Frogel1978}, \citet{Bessell1998} and \citet{Elias1982}. A description of the SMARTS data reduction can be found in \citet{Bonning2012}.

The Wide-field Infrared Survey Explorer \citep[WISE;][]{Wright2010} is an all-sky survey mission in the mid-infrared that operated between 2009 January and 2010 October. It took images  that were 47 arcminutes in width, every 11 seconds. It was capable of imaging near-infrared (3.4 and 4.6\,$\mu$m) and mid-infrared (12 and 22\,$\mu$m) bands. WISE made an observation of PKS\,2326$-$502 on 2010 July 25 that was not simultaneous with flare A, flare B or period Q. However, since the source seemed to be in a low $\g$-ray state at that time, we tentatively include the data as part of period Q. This observation was made at all four wavebands. The data was drawn from the WISE Preliminary Data Release\footnote{http://wise2.ipac.caltech.edu/docs/release/prelim/}. The fluxes were 1.36$\pm$0.03 mJy (13.42$\pm$0.03), 2.26$\pm$0.05 mJy (12.19$\pm$0.03), 7.88$\pm$0.17 mJy (9.00$\pm$0.02), 21.67$\pm$0.98 mJy (6.45$\pm$0.05).

\section{Radio Observations}\label{radio}

PKS\,2326$-$502  is observed by the Australia Telescope Compact Array (ATCA) at several radio frequencies as part of the TANAMI blazar monitoring program \citep{Stevens2012}. Data from this monitoring were available quasi-simultaneously with the quiescent state (2012 January 15) and the 2012 flare (2012 June 29). No ATCA data were available during the 2010 flare. The ATCA is an array consisting of $6\times22$\,m radio antennas with adjustable baselines and a longest baseline of 6 km. The array configuration is changed every few weeks. However, as PKS\,2326$-$502 is a point source even for ATCA's longest baseline, ATCA's configuration does not affect our observations. The receivers at ATCA can be quickly changed allowing observations to be made over a large range of frequencies in a short period of time. The array is located in northern New South Wales, at a latitude of $-$30$^\circ$ and altitude 237\,m above sea level. During the quiescent state ATCA observed at 5, 9, 17, 19, 38 and 40\,GHz. The 2012 flare has observations at 9, 17, 19, 38 and 40\,GHz. Snapshot observations of PKS 2326$-$502 of several minutes duration were made at each frequency, and calibrated against the ATCA primary flux calibrator PKS\,1934$-$638. Observations at 38/40 GHz are preceded by a scan on a bright nearby AGN to apply corrections to the global pointing model. Data reduction was carried out in the standard manner with the miriad software package\footnote{http://www.atnf.csiro.au/computing/software/miriad/}.



\section{Results}\label{results}

Using the data described above, MWL SEDs were constructed for both flaring states and the quiescent state. 
Then we modeled the three states with a one-zone leptonic model, including
synchrotron, synchrotron self-Compton (SSC), and external Compton
(EC).  \citet{finke08_SSC} and \citet{Dermer2009} have more details on the modeling,
 describing an approach to modeling using the radio to X-ray flux to obtain 
an electron spectrum. That spectrum is then used to deduce the synchrotron self Compton spectrum in the the Thompson through Klein-Nishina regions.
The external isotropic radiation field was assumed to be
monochromatic and isotropic in the frame of the host galaxy and black
hole.  \object{PKS 2326$-$502} has an uncertain redshift of $z=0.518$
based on a weak detection of a single emission line, \ion{Mg}{2}
$\lambda2798$ \citep{Jauncey1984}.  We use this redshift for our
modeling, but one should keep its uncertainty in mind.  The size
of the emitting region was constrained by the variability doubling timescale,
found to be about a day from the $\gamma$-ray light curve (Figure \ref{gammalightcurve}).

\begin{figure}
\vspace{2.2mm} 
\epsscale{1.0} 
\plotone{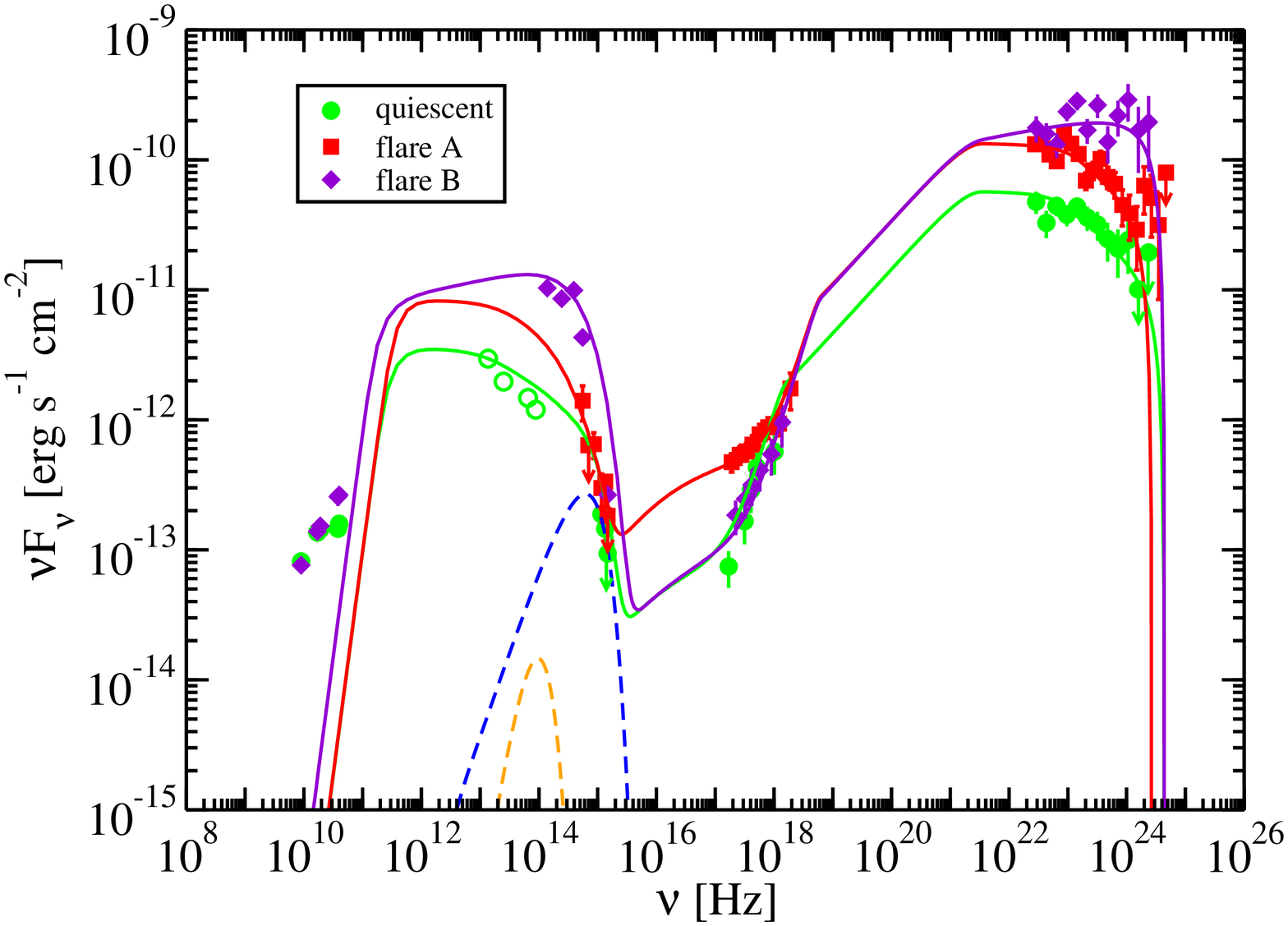}
\caption{Broadband SED model for PKS\,2326$-$502. The data are modeled with a one-zone leptonic model of blazar emission \citep{finke08_SSC}. The low-energy component is modeled as a combination of synchrotron emission from the jet and thermal emission from the accretion disk. A double broken power law is used to model the electron distribution during the quiescent state flare A. A change in the size of the emitting region is used to model flare B. The smaller dashed green curve represents emission from the dust torus and the larger dashed blue curve is accretion disk emission. The high-energy component is explained by inverse Compton scattering of dust torus photons by electrons within the jet. Hollow circles represent non-simultaneous data.
}
\label{SED_fig1}
\vspace{2.2mm}
\end{figure}

The modeling results are presented in Figure \ref{SED_fig1} and Table
\ref{table_fit1}.  A detailed explanation of the model parameters can
be found in \citet{Dermer2009}.  We chose a relatively weak accretion
disk \citep{Shakura1973} to model the optical portion of the SED in the quiescent
state, which is consistent with the lack of features in the source's
optical spectrum \citep{Jauncey1984}.  The black hole's mass is not
known, but we chose a standard value of $10^9\ M_\odot$, which is
consistent with the optical SED. 
We found that all three states could
be modeled with a broken power-law electron distribution.  However, the SED in the $10^{19}-10^{21}$ Hz (40 keV$-$4 MeV)
range does not very closely resemble the SED of other FSRQs.  Therefore, we
chose to model the quiescent and the flare A SED with a double broken
power-law.  In this case, its electron distribution is
\begin{equation}
N_e(\g) \propto \left\{ \begin{array}{ll}
\g^{\prime -p_1} & \gp_1 < \gp < \gp_{brk1} \\ 
\g^{\prime -p_2} & \gp_{brk1} < \gp < \gp_{brk2} \\
\g^{\prime -p_3} & \gp_{brk2} < \gp < \gp_2
\end{array}
\right. 
\end{equation}
where $\gp$ is the electron Lorentz factor in the frame co-moving with
the blob.  This type of electron distribution is not without
precedence, e.g. a double broken power-law electron distribution was used
by \citet{dammando13} to model the SED of PKS\,0537$-$441.  Here, it also
allowed us to obtain models closer to equipartition.  This electron
distribution could be tested by observations by {\em NuSTAR}, which
would constrain the energy range that necessitates the double broken
power-law. Flare B required only a broken power law electron distribution to explain the X-ray to $\g$-ray portion of the SED.


We found that the quiescent state and flare A could be modeled
by only varying the electron distribution between states.  This is
similar to the flaring states found in PKS\,0537$-$441
\citep{dammando13}, `flare B' from PKS\,2142$-$75
\citep{dutka13_2142}, and the flaring states of 4C+21.35
\citep{ackermann14_4c21.35}.  Modeling flare B
by only varying the electron distribution would result in the SSC
emission over-producing X-rays.
Therefore, the variation
in another parameter is necessary.  We chose to change the variability
timescale $t_v$, which has the effect of expanding the size of the
emitting region in the model ($R^\prime_b$).  The larger variability
timescale is still consistent with the light curve presented in Figure \ref{gammalightcurve}.  

According to the relation from \citet{ghisellini08}, the broad-line region
(BLR) radius, $R_{BLR}$ (where $R_{BLR,17}$ is $R_{BLR}$ in units of 10$^{17}$ cm) is given by, 
\begin{eqnarray}
R_{BLR,17} = L_{disk,45}^{1/2} 
\end{eqnarray}
where $L_{disk,45}$ is the disk luminosity times 10$^{45}$. In this case, 
$R_{BLR} = 5.5\times10^{16}\ \cm$.  Here we use the notation
$A = 10^x A_x$ and cgs units.  This value would not be consistent with
the size of the emitting region inferred from the variability
timescale, although we note that size is only a soft upper limit based on the most rapid changes seen in the lightcurve, not the most rapid possible changes.
Also, the disk luminosity is not well-constrained from the SED.
We still chose to model the source with the dust torus as the source of seed photons, although one should keep the caveats in
mind.  The energy density from the dust torus with temperature $T$
which reprocesses a fraction $\xi$ of the disk luminosity, assuming
the radiation is dominated by the inner dust radius, is
\begin{eqnarray}
u_{dust} = 2.4\times10^{-5} \xi_{dust,-1} T_3^{5.2}\ \erg\ \cm^{-3}
\end{eqnarray}
\citep{nenkova08,sikora09}.  The model has dimensionless
seed photon energy and energy density $\e_{seed}=9.0\times10^{-7}$ and
$u_{seed}=1.8\times10^{-4}\ \erg\ \cm^{-3}$, respectively.  This
implies $T_3=1.8$ (in units of 10$^{3}$ K), $\xi_{-1} = 0.35$ giving
$L_{dust}=1.1\times10^{43}$ erg s$^{-1}$, and $R_{dust}=5.6\times10^{17}\ \cm$.
Assuming a conical jet, the jet half-opening angle must be greater
than $\alpha=R_b^\prime/R_{dust} = 2.8\arcdeg$, which is consistent with measurements of other jet opening angles from VLBI \citep{Jorstad2005}.

The accretion power in this model is $P_{acc}=L_{disk}/\eta =
3.6\times10^{45}\ \erg\ \s^{-1}$.  The models for the three states
give results that imply the electron energy density is almost in
equipartition with the magnetic energy density.  The total jet powers,
$P_j=P_{j,e}+P_{j,B}$ make up a large fraction of the accretion
powers, ranging from $P_j/P_{acc} = 0.72$ for flare B to
$P_j/P_{acc}= 1.5$ for flare A.  The jet seems to be
highly efficient for this source, with the jet power possibly even exceeding the power
from accretion only.  This may be possible in magnetically arrested
accretion onto a black hole with nearly maximal spin \citep{tchek11}.
However, there are large uncertainties in the jet power from the
modeling, and the disk luminosity is not well-constrained.  Better
data in the optical band, especially a better optical spectrum, could
constrain the disk luminosity better.


\begin{sidewaystable*}
\footnotesize
\begin{center}
\caption{Model parameters for the SED shown in Fig.~\ref{SED_fig1}.
\label{table_fit1}}
\begin{tabular}{lcccc}
Parameter & Symbol & Quiescent & flare A & flare B \\
\hline
Redshift & 	$z$	& 0.518	& 0.518	& 0.518	\\
Bulk Lorentz Factor & $\Gamma$	& 30	& 30 & 30	  \\
Doppler factor & $\delta_D$	& 30	& 30 & 30	  \\
Magnetic Field [G]& $B$         & 0.50 & 0.50 & 0.50 \\
Variability Timescale [s]& $t_v$       & $4.6\times10^4$ & $4.6\times10^4$ & $1.3\times10^5$ \\
Comoving radius of blob [cm]& $R^{\prime}_b$ & 2.7$\times$10$^{16}$ & 2.7$\times$10$^{16}$ & $7.7\times10^{16}$ \\
\hline
Electron Spectral Index 1 & $p_1$       & 2.0 & 2.0 & 2.0 \\
Electron Spectral Index 2 & $p_2$       & 3.0 & 3.0 & 2.8 \\
Electron Spectral Index 3 & $p_3$       & 3.6 & 3.6 & n/a \\
Minimum Electron Lorentz Factor & $\gamma^{\prime}_{min}$  & $2.0$ & $4.0$ & $4.0$ \\
Break Electron Lorentz Factor 1 & $\gamma^{\prime}_{brk1}$ & $1.0\times10^2$ & $1.0\times10^2$ & $1.0\times10^2$ \\
Break Electron Lorentz Factor 2 & $\gamma^{\prime}_{brk2}$ & $5.1\times10^2$ & $6.7\times10^3$ & n/a  \\
Maximum Electron Lorentz Factor & $\gamma^{\prime}_{max}$  & $4.0\times10^3$ & $3.0\times10^3$ & $4.0\times10^3$  \\
\hline
Black hole Mass [$M_\odot]$ & $M_{BH}$ & $1.0\times10^9$ & $1.0\times10^9$ & $1.0\times10^9$ \\
Disk luminosity [$\erg\ \s^{-1}$] & $L_{disk}$ & $3.0\times10^{44}$ & $3.0\times10^{44}$ & $3.0\times10^{44}$ \\
Inner disk radius [$R_g$] & $R_{in}$ & $6.0$ & $6.0$ & $6.0$ \\
Accretion Efficiency & $\eta$ & 1/12 & 1/12 & 1/12 \\
Seed photon source energy density [$\erg\ \cm^{-3}$] & $u_{seed}$ & $1.8\times10^{-4}$ & $1.8\times10^{-4}$ & $1.8\times10^{-4}$ \\
Seed photon source photon energy & $\e_{seed}$ & $9.0\times10^{-7}$ & $9.0\times10^{-7}$ & $9.0\times10^{-7}$ \\
Dust Torus luminosity [$\erg\ \s^{-1}$] & $L_{dust}$ & $1.1\times10^{43}$ & $1.1\times10^{43}$ & $1.1\times10^{43}$ \\
Dust Torus radius [cm] & $R_{dust}$ & $5.6\times10^{17}$ & $5.6\times10^{17}$ & $5.6\times10^{17}$ \\
\hline
Jet Power in Magnetic Field [$\erg\ \s^{-1}$] & $P_{j,B}$ & $1.3\times10^{45}$ & $1.3\times10^{45}$  & $1.0\times10^{45}$ \\
Jet Power in Electrons [$\erg\ \s^{-1}$] & $P_{j,e}$ & $2.1\times10^{45}$ & $4.1\times10^{45}$ & $1.6\times10^{45}$ \\
\hline
\end{tabular}
\end{center}
\end{sidewaystable*}

\section{Discussions and Conclusions}\label{conclusions}

In order to study the origins of the high-energy emission, observations and archival data from \textit{Fermi}, \textit{Swift}, SMARTS, REM, WISE, and ATCA were used to construct the MWL SEDs of PKS\,2326$-$502 during a quiet and two flaring $\gamma$-ray states. Period Q was a period of `average' $\gamma$-ray activity for PKS\,2326$-$502 and was observed to provide a baseline to compare flare A and flare B against. A one-zone leptonic model can appropriately describe the SED constrained by this data.

Modeling flare A and flare B required different changes to the parameters of the SED model that describes the state Q. Flare A only required changing the electron distribution.  However, the increased emission during flare B could not be explained by changes in the electron distribution alone and required a change in the size of the emitting region as well. This fits with a previous classification scheme for blazar flares \citep{dutka13_2142}. Within this scheme AGN exhibit flares of two types. Type 1 flares (like flare A) are those that show changes in the SED from a quiescent state which can be explained entirely by modifying the electron distribution. Flare B is a type 2 flare. These require a change in the electron distribution but that is not sufficient; a change to either the magnetic field or the size of the emitting region must be made to match the emission across the SED. As is seen here with PKS\,2326$-$502  both types of flares can occur from the same source. These classes can each be divided into subclasses. Type 1a shows increased emission at both the optical and $\gamma$-ray wavelengths, type 1b only shows an increase in the $\gamma$-ray band. X-ray emission can see either an increase or not for type 1 flares. Type 2a features flaring in the optical and $\gamma$-ray but not in the X-ray, type 2b displays flaring in optical, X-ray and $\gamma$-ray. Here we would classify flare A as a type 1a and flare B as a type 2b flare. Flare A was very different from flare B in that it was also the beginning of a long, sustained period of high flux whereas flare B showed a very sharp peak and a faster return to average fluxes.

This emerging classification scheme may allow us to gain additional physical insight into the processes that cause blazar flares.  Our models suggest that these processes  are not uniform, and that they can arise from different physical conditions.  This sort of behavior could be expected from a turbulent, outflowing plasma.  Type 1 flares, where changes to the electron distribution are sufficient to cause the flare, may result from electrons moving into or out of the emitting region. Or, the electrons within the emitting region could experience a bulk acceleration.  Type 2 flares could be explained by shocks changing the shape of the emitting region.  A compression, or expansion, of the emitting region could have an effect on the magnetic energy density.  We may be able to explain flares as a purely magnetic phenomenon as well.  The diversity in the behavior of simultaneous SEDs may allow us to probe the behavior of jets at scales which are too small to resolve with current observational techniques.

We expect to improve on this classification scheme and to gain new insights into the high-energy emission processes in PKS\,2326$-$502  through observations of additional active states. VLBI observations of PKS\,2326$-$502  are being made by the TANAMI program using the Long Baseline Array based in Australia. These will allow us to determine the jet kinematics, including a more direct measure of the Doppler factor. VLBI monitoring can show us the emergence of new jet components. The emergence of jet components is usually correlated with $\gamma$-ray flares \citep{Marscher2012}: broadband observations of flares combined with VLBI monitoring could determine if new components emerge with specific types of flares. ALMA observations will constrain the sub-mm region of future SEDs leading to a much better determination of the synchrotron peak. Observations with \textit{NuSTAR} in the hard X-ray regime would help constrain the SED in the region where the power law breaks occur. More broadband, quasi-simultaneous observations of other AGN are essential to improving and verifying the tentative classification scheme outlined above, thus helping solve the puzzle of high-energy emission in blazars.

\acknowledgments
\section{Acknowledgements}

This research was funded in part by NASA through {\em Fermi} Guest
Investigator grants NNH09ZDA001N, NNH10ZDA001N, and NNH12ZDA001N. This
research was supported by an appointment to the NASA Postdoctoral
Program at the Goddard Space Flight Center, administered by Oak Ridge
Associated Universities through a contract with NASA.  This
publication makes use of data products from the Wide-field Infrared
Survey Explorer, which is a joint project of the University of
California, Los Angeles, and the Jet Propulsion Laboratory/California
Institute of Technology, funded by the National Aeronautics and Space
Administration.  The Australia Telescope Compact Array is part of the
Australia Telescope National Facility which is funded by the
Commonwealth of Australia for operation as a National Facility managed
by CSIRO.  This research has made use of data from the NASA/IPAC
Extragalactic Database (NED), operated by the Jet Propulsion
Laboratory, California Institute of Technology, under contract with
the National Aeronautics and Space Administration; and the SIMBAD
database (operated at CDS, Strasbourg, France). This research has made
use of NASA's Astrophysics Data System. This research has made use of
the United States Naval Observatory (USNO) Radio Reference Frame Image
Database (RRFID).  This paper has made use of up-to-date SMARTS
optical/near-infrared light curves that are available at
\url{http://www.astro.yale.edu/smarts/glast/home.php}. F. K. acknowledges funding from the European UnionÕs Horizon 2020 research and innovation programme under grant agreement No 653477.

The {\em Fermi} LAT Collaboration acknowledges generous ongoing support
from a number of agencies and institutes that have supported both the
development and the operation of the LAT as well as scientific data analysis.
These include the National Aeronautics and Space Administration and the
Department of Energy in the United States, the Commissariat \`a l'Energie Atomique
and the Centre National de la Recherche Scientifique / Institut National de Physique
Nucl\'eaire et de Physique des Particules in France, the Agenzia Spaziale Italiana
and the Istituto Nazionale di Fisica Nucleare in Italy, the Ministry of Education,
Culture, Sports, Science and Technology (MEXT), High Energy Accelerator Research
Organization (KEK) and Japan Aerospace Exploration Agency (JAXA) in Japan, and
the K.~A.~Wallenberg Foundation, the Swedish Research Council and the
Swedish National Space Board in Sweden. Additional support for science analysis during the operations phase is gratefully
acknowledged from the Istituto Nazionale di Astrofisica in Italy and the Centre National d'\'Etudes Spatiales in France.

{\it Facilities:} \facility{ATCA}, \facility{Fermi}, \facility{Swift}, \facility{SMARTS}, \facility{WISE}.



\bibliographystyle{apj}
\bibliography{aa_abbrv,mnemonic,tanami}

\clearpage

\clearpage

\clearpage

\end{document}